\documentstyle[12pt,aaspp]{article}
\articleid{11}{14}
\begin{document}

{\obeylines
\hfill{YITP/U-94-30}
\hfill{IUCAA - 38/94}}

\title{ SKEWNESS OF COSMIC MICROWAVE BACKGROUND TEMPERATURE
FLUCTUATIONS DUE TO NON-LINEAR GRAVITATIONAL INSTABILITY }

\author {Dipak Munshi, Tarun Souradeep}
\affil {Inter-University Center for Astronomy and Astrophysics, \\
Post Bag 4, Ganeshkhind, Pune 411007, India \\ munshi@iucaa.ernet.in
\\ tarun@iucaa.ernet.in }
\and
\author {Alexei A. Starobinsky}
\affil {Yukawa Institute for Theoretical Physics, Kyoto University,
Uji 611, Japan \\ and \\
Landau Institute for Theoretical Physics, Kosygina St. 2, Moscow 117334,
Russia \\ alstar@landau.ac.ru}

\begin{abstract}
Skewness of temperature fluctuations of the cosmic microwave background
(CMB) produced by initially Gaussian adiabatic perturbations with
the flat (Harrison-Zeldovich) spectrum, which arises due to non-linear
corrections to a gravitational potential at the matter-dominated
stage, is calculated quantitatively. For the standard CDM model,
the effect appears to be smaller
than expected previously and lies below the cosmic variance limit even
for small angles. The sign of the skewness is opposite to that of the
skewness of density perturbations.
\end{abstract}

\keywords { cosmology: theory --- cosmic microwave background  ---
large-scale structure of the Universe}

\vfill
\eject

\section{Introduction}

The reliable detection of fluctuations of the CMB temperature
$\Delta T (\theta, \varphi)/T$ at large angular scales by the COBE
group (with the data being in a very good agreement with a
prediction made on the basis of the inflationary scenario of
the early Universe 10 years before the detection)
stimulates further investigation of subtler effects. Among the most
important of them are possible deviations of the statistics of these
fluctuations from the Gaussian one. The basic result which follows from
all sufficiently simple
variants of the inflationary scenario (contrary to that of rival
theories based on topological defects such as cosmic strings, etc.) is
that the statistics of the $\Delta T/T$ fluctuations
is Gaussian because they are linearly connected to quantum
vacuum fluctuations of a very weakly interacting scalar field (the
inflaton). Thus, in the leading (linear) approximation,
the mean CMB skewness $C_3(0)=\langle (\Delta T/T)^3 \rangle =0$
where $\langle \rangle$ denotes averaging with respect
to different realizations of stochastic space-time metric
perturbations of the Friedmann-Robertson-Walker cosmological model
which produce $\Delta T/T$.
Note that the corresponding observable
quantity is $\bar C_3(0)=(4\pi)^{-1}\int \left(\Delta T(\theta,
\varphi)/T\right)^3~d\Omega$,
i.e., the average over the sky of one particular realization.
Therefore, $\bar C_3(0)$ generally differs from $C_3(0)$ by
a so-called cosmic variance which is of the order of
$\left(\langle (\Delta T/T)^2\rangle \right)^{3/2}$ (see,
e.g., Srednicki 1993).
In practice, even $\bar C_3(0)$ is unachievable because of a finite
beam width of antennas, incomplete sky coverage, etc.

There exist different physical effects
which may result in the appearance of a small, but non-zero mean
skewness. Some of them are connected with non-linear corrections to
the initial spectrum of scalar (adiabatic) metric perturbations
which were generated during inflation (Falk et al. 1993, Gangui
et al. 1994), the corresponding part of the CMB mean
skewness may be called primordial. Other effects which take place after
recombination (Luo \& Schramm 1993) produce a secondary skewness.
In this paper, we will concentrate on a detailed calculation
of the secondary skewness produced by non-linear corrections
to the primordial gravitational potential $\Phi$ which arises
due to the same gravitational instability at the matter-dominated
stage that leads to formation of galaxies and the large-scale
structure of the Universe. The corresponding contribution to
$\Delta T/T$ is contained in the non-local term of the Sachs-
Wolfe expression for $\Delta T/T$ (Sachs \& Wolfe 1967), it is also
called the Rees-Sciama effect (Rees \& Sciama 1968, a different view
on this effect is presented in Zeldovich \& Sazhin 1987).

The reason for our primary interest in this effect is that its
contribution to $\Delta T/T$, though formally being of  second
order in powers of a small initial gravitational
potential $\Phi(0)=\phi_0({\bf r})$, is not much less than the main
linear effect. Its value may be estimated as $\Delta T/T \sim
\Phi \delta \rho /\rho$ where $\delta \rho /\rho$
is the present {\it rms} density perturbation inside some
linear characteristic size $L$. For $L \sim R_{eq}\approx 30h^{-1}$ Mpc
corresponding to the angular scale $\vartheta \sim LH_0/c \sim 30'$ (if
viewed from the redshift $z\sim 1$),
it is $\sim \phi_0^2 z_{eq}$
where $z_{eq}\approx 4\cdot 10^4 h^2 \kappa^{-1}$
is the redshift of matter-radiation energy equality; therefore,
it may reach $10^{-6}$ (Martinez-Gonzales, Sanz \& Silk 1992). Here
$h$ is the value of the Hubble constant $H_0$ in units of $100$ km/s/Mpc
and $\kappa =\epsilon_{rad}/\epsilon_{\gamma}=1.68$ for 3 types of light
($m\ll 1$ eV) neutrinos with standard concentrations.

On the basis of the above argument, it was  conjectured (Luo \& Schramm 1993)
that the secondary
CMB skewness produced by the non-local part of the Sachs-Wolfe effect
will dominate the primordial skewness generated in the inflationary scenario.
However, our calculations presented in the next section
show that this is not the case for the standard CDM model. The
secondary skewness appears to be small and of the order of primordial
skewness. A possibility of getting a larger skewness without introducing
late-time phase transitions is discussed in the Sec. 3. The section also
contains our conclusions, as well as comparison with the cosmic variance
of the skewness in the case of purely Gaussian perturbations. Details of
our calculations are displayed in Appendices A and B.

\section{Method of calculation}

For large angular scales $\vartheta \ge 2^{\circ}$, CMB temperature
fluctuations produced by adiabatic perturbations are given by the
Sachs-Wolfe formula:
\begin{eqnarray}
{\Delta T \over T}(\theta, \varphi)  &=& \left({\Delta T\over T}
\right)_{loc}+\left({\Delta T\over T}\right)_{non-loc}=
{1\over 3} \Phi (r_0,\theta, \varphi) + 2\int_{\eta_{rec}}^{\eta_0}
\left({\partial \Phi (\eta ,{\bf r})\over \partial \eta}\right)_
{r=\eta_0-\eta}d\eta~, \nonumber \\
{\bf r} &=& (r, \theta, \varphi),~~~r_0=\eta_0-\eta_{rec},~~~ \eta =\int
{dt\over a(t)}~, \label{SW}
\end{eqnarray}
where $a(t)$ is the scale factor of the FRW model, $\eta_0$ is the
present conformal time and $\eta_{rec}$ is the conformal recombination
time, $\eta_0/\eta_{rec} \approx z_{rec}^{1/2}\sim 30$. More exact
calculation for the CDM model with 3 light neutrino species and
$z_{rec}=1100$ gives $\eta_0/\eta_{rec}=49.6$.  At the
matter-dominated stage in the absence
of  spatial curvature and the cosmological constant,
$a(t)\propto t^{2/3},~\eta = 3t/a(t)$. For smaller angles
$5' <\vartheta <2^{\circ}$ where  standard recombination may be
still considered as instantaneous, additional local terms in
Eq. (\ref{SW}) appear which describe the Doppler and the Silk
effects (see, e.g., Starobinsky 1988), but the non-local term
remains the same. In the linear approximation, this term is exactly
zero for any power-law $a(t)$, though it may contribute
significantly if $a(t)$ deviates from a power-law behaviour after
recombination, e.g., due to the cosmological constant
(Kofman \& Starobinsky 1985), decaying relativistic particles
(Kofman, Pogosyan \& Starobinsky 1986) or  spatial curvature
(Wilson 1983, Abbott \& Schaeffer 1986). However, for the purely
matter-dominated stage in the flat Universe that we consider,
the non-local term produces a non-zero contribution to $\Delta T/T$
if non-linear corrections to the gravitational
potential $\Phi$, arising due to  gravitational instability
in the Universe (see, e.g., Peebles 1980), are taken into account.
These corrections are also responsible for the appearance of
a non-zero skewness of CMB in the case of a Gaussian initial (linear)
potential.

So, we expand the (peculiar) gravitational potential into a series
in powers of a density enhancement: $\Phi ({\bf r},t)=
\Phi^{(1)}+\Phi^{(2)}+...$. The linear term $\Phi^{(1)}=
\phi_0({\bf r})$ is assumed to be Gaussian. Further, we consider
the case of a flat (Harrison-Zeldovich) initial spectrum.
Then the two-point correlation function of the linear potential
is given by
\begin{equation}
\xi_{\phi}(r)=\langle \phi_0(0) \phi_0(r) \rangle =
\int _0^{\infty}P_{\phi}(k){\sin kr\over kr}k^2dk =
B\int c^2(k){\sin kr\over kr}{dk\over k}~,  \label{cor}
\end{equation}
where $c(k)$ is the standard transfer function of the CDM model.
For physical scales $R\equiv ra(t_0)\gg R_{eq}$,
$\xi_{\phi}(r)\approx \xi_{\phi}(0)-B\ln(r/r_{eq})$ (strictly speaking,
the constant $\xi_{\phi}(0)$ is infinite but it is unobservable, so no
difficulties arise). The constant $B$ is related to the
amplitude normalization $A$
introduced in Starobinsky 1983 by the relation $B=9A^2/200\pi^2$.
The {\it rms} CMB quadrupole value is expressed through it by
$Q^2_{rms-PS}/T^2= 5B/108$, so that $B=1.16\times 10^{-9}$
for the presently preferred values $T=2.726~K$  and
$Q_{rms-PS}=20~\mu K $ (see, e.g., Gorski et al. 1994).

The second-order term is given by (Peebles 1980, notations of the
paper by Munshi \& Starobinsky 1994 are used below):
\begin{eqnarray}
\Phi^{(2)}={\eta^2\over 42} (5\triangle^{-1}P+Q);~~~~~~~\nonumber \\
P(\vec r)=(\triangle \phi_0)^2+\nabla \phi_0 \nabla (\triangle \phi_0)
=\nabla (\nabla \phi_0 \triangle \phi_0),~~~
Q(\vec r)=(\nabla \phi_0)^2. \label{nl}
\end{eqnarray}
To get a finite expression for $C_3(0)$, it is sufficient to use the
``renormalized'' quantity:
\begin{equation}
\Phi_{ren}^{(2)}({\bf r}, \eta )=\Phi^{(2)}({\bf r}, \eta )-
\langle \Phi^{(2)}({\bf r}, \eta )\rangle~, \label{ren}
\end{equation}
(the last term in (\ref{ren}) depends on $\eta$ only) that corresponds
to the subtraction of an unobservable constant
(monopole) term from $\Delta T/T$. More rigorous justification of
this prescription follows from the fact that the difference $\Phi
({\bf r}, \eta )-\Phi({\bf r}_0, \eta )$, where ${\bf r}_0$ denotes
an observer (e.g., our) location, is an observable and finite
quantity both in linear and non-linear regimes.
Further, we omit the subscript ``ren''. Note that the subtraction
(\ref{ren}) does not remove the whole contribution to $C_3(0)$
from monopole terms, in each of $\Delta T/T$ some finite part remains.

A non-zero contribution to the CMB mean skewness in the lowest
(fourth) order in $\Phi$ is given by
\begin{equation}
C_3^{(4)}(0)=3\langle \left({\Delta T\over T}\right)_{loc}^2
\left({\Delta T\over T}\right)_{non-loc}\rangle~. \label{four}
\end{equation}
Let us now remove a monopole component from each of ${\Delta T\over T}$
completely:
\begin{equation}
\left({\Delta T\over T}\right)_S={\Delta T\over T}-{1\over 4\pi}
\int ~{\Delta T \over T}~d\Omega~.  \label{mon}
\end{equation}
Then the term (\ref{four}) can be represented as
\begin{equation}
C_3^{(4)}(0)=I_{UUU}-I_{UUM}-2I_{UMU}+2I_{UMM}+I_{MMU}-I_{MMM}
\label{sub}
\end{equation}
where the subscript $U$ means substitution of an unsubtracted
${\Delta T\over T}$ into (\ref{four}) and $M$ - substitution of
a monopole term there (the gravitational potential in both terms is assumed
to be renormalized according to (\ref{ren})). Strictly speaking,
a dipole component should be subtracted from $\left({\Delta T\over T}
\right)_S$
further, but this correction appears to be smaller than the correction
due to the monopole subtraction (\ref{mon}) (e.g., $I_{UUD}\approx
0.74~I_{UUM}$, see Appendix A) and practically does
not change the final result for $C_3(0)$.

The main term in the right-hand side of Eq. (\ref{sub}) is
 $I_{UUU}$. After substitution of (\ref{nl}) into Eq. (\ref{SW}),
it takes the form
\begin{equation}
I_{UUU}={4\over 63}\int_{\eta_{rec}}^{\eta_0}\eta \left(\xi_{\phi}'^2+
5\triangle^{-1}((\triangle \xi_{\phi})^2+ \xi_{\phi}'(\triangle
\xi_{\phi})'\right)_{r=\eta -\eta_{rec}}\, d\eta \label{UUU}
\end{equation}
where $\xi_{\phi}$ is given in Eq. (\ref{cor}) and the prime means
derivative with respect to $r$. Note that for $r\gg r_{eq}$,
$\triangle^{-1}\left((\triangle \xi_{\phi})^2+\xi_{\phi}'
(\triangle \xi_{\phi})'\right)=-{B^2\over 2r^2}$. Thus, if we are
speaking about the skewness of temperature fluctuations smoothed
over a scale $R_s>R_{rec}=a(\eta_0)\eta_{rec}$ (e.g., due to a finite
antenna beam width), then the integral in (\ref{UUU})
diverges at both large and small $\eta$ logarithmically:
$I_{UUU}=-{2\over 21} B^2\ln {R_h\over R_s}$, where
$R_h=a(\eta_0)\eta_0$ is a present cosmological horizon of the FRW
model. This means that the growth of $\left({\Delta T
\over T}\right)_{non-loc}$ in (\ref{four}) at late times due to
$\Phi^{(2)}$ is exactly cancelled by decay of correlations
between $\left({\Delta T\over T}\right)_{loc}$ and $\left({\Delta T
\over T}\right)_{non-loc}$ at large spatial separations. Another
conclusion following from the behaviour of the integrand in Eq.(\ref{UUU})
is that the contribution of the small redshift region $z\sim 1,~\eta
\sim \eta_0$ to the skewness is subdominant, it is smaller than the
total effect in a few ($\sim 1/\ln(\eta_0/\eta_{rec})$) times. That is
why higher-order non-linear corrections do not noticeably change the
fourth order result for the skewness (we check it explicitly
for the leading sixth order term, see Appendix B).  Clearly, the same
statement is true for the case of a blue-tilted initial spectrum $n>1$.
However, in the opposite case of a significantly red-tilted spectrum
($n \stackrel {<}{\sim} 0.9$) higher-order non-linear corrections
may become crucial since the main effect originates at recent times.

In terms of the multipole decomposition,
this behaviour corresponds to equal contributions to $C_3(0)$ from each
logarithmic interval of $l$ for $l\gg 1$. In other words, if
$C_3(0)=\sum_l C_l$, then $C_l =-{2B^2\over 21(l+0.5)}$ for $l\gg 1$. Using
the monopole value $C_0$ obtained below (see Appendix A and Eq. (\ref{C3})),
this fit may be made better by changing $(l+0.5)$ to $(l+1/3)$. So, if we take
the smoothing angle $\theta_{FWHP}=10^{\circ}$ as in the smoothed COBE maps
($\theta_g=0.425\theta_{FWHP}=1/13.5$), then
\begin{equation}
C_3(10^{\circ})=-0.16B^2~, ~~~~\langle \left(\Delta T(10^{\circ})\right)^3
\rangle =4.4~(\mu K)^3 \label{sm}
\end{equation}
(with monopole and dipole terms subtracted).
Since  $C_2(10^{\circ})=0.18B$, the smoothed large-angle skewness parameter
$S_3(10^{\circ})\equiv {C_3(10^{\circ})\over C_2^2(10^{\circ})} \approx -5$.

For the unsmoothed distribution of temperature fluctuations,
the main contribution to $I_{UUU}$ comes from the vicinity of the
recombination surface $\eta - \eta_{rec} \sim \eta_{eq}$ as a result
of properties of the transfer function $c(k)$. However, this distance
is still significantly larger than the thickness of the recombination
surface. Hence, the latter may be neglected for our problem. The
corresponding angular range is $10'-30'$. Here lies the first
acoustic, or Doppler peak, so the first term alone in the right-hand side of
Eq. (\ref{SW}) should be corrected by  accounting for the Doppler
and the Silk effects (but not the second one). As a result, we obtain
(details of the calculation are presented in the Appendix A):
\begin{equation}
I_{UUU} \approx - 2.2~B^2.
\label {IUUU}
\end{equation}
Other terms in the right-hand side of (\ref{sub}) may be calculated
in the approximation $\eta_{rec}=0,~c(k)=1$. From symmetry
considerations, $I_{UMM}=I_{MMU}=I_{MMM}$. It is shown in the Appendix
A that $I_{UUM}=-{2\over 21}({\pi^2\over 8}-\ln 2)B^2\approx -0.05
B^2$. Other auxiliary integrals entering into Eq. (\ref{sub}) are also given
there. The dipole contribution can be estimated to be $\sim -0.07 B^2$
using  the abovementioned fit $C_l =-{2B^2\over 21(l+ 1/3)}$.
As a result,

\begin{eqnarray}
C_3^{(4)}(0) &=& I_{UUU}-I_{UUM}-2(I_{UMU}-I_{UMM}) - C_1\nonumber \\
&=& -2.2B^2 + 0.05B^2+ 0.24B^2 + 0.07 B^2\approx  -1.8 B^2~, \nonumber \\
\langle \left(\Delta T\right)^3\rangle &\approx& 50~(\mu K)^3.
 \label{C3}
\end{eqnarray}

With a good accuracy,
$\sigma_T^2\equiv C_2(0)=\langle \left({\Delta T\over T}\right)^2
\rangle \approx 10^{-9} \approx B$ where
$C_2(\vartheta)\equiv \langle {\Delta T\over T}(0)~
{\Delta T\over T}(\vartheta)\rangle$ is the 2-point angular
temperature correlation function. Therefore, the skewness parameter
$S_3(0)\equiv {C_3(0)\over C_2^2(0)} \sim -2$
is of the same order as the primordial skewness
considered in Gangui et al. 1994. The unsmoothed value of the skewness
$C_3(0)$ is $\sim 11$ times more than $C_3(10^{\circ})$ because
the main effect comes from the angles $10'-30'$. However, the
unsmoothed skewness parameter $S_3(0)$ is smaller than $S_3(10^{\circ})$
because fluctuations themselves (i.e. $C_2$) are significantly larger at
small angles.

Note the negative sign of $S_3$.
Its physical explanation is that it reflects the existence of large
regions with $\Phi$ positive and growing with time (corresponding
to voids) which produce positive $\left({\Delta T\over T}
\right)_{non-loc}$
and relatively small regions of negative and decreasing $\Phi$
which produce smaller cold spots but with larger absolute values of
${\Delta T\over T}$. It is these cold spots that make the main
contribution to the skewness and determine its sign. The sign is
opposite to the sign of a skewness of density perturbations
${\delta \rho \over \rho}$. In the latter case, the main contribution
to the skewness is produced by regions with ${\delta \rho \over \rho}
>0$ which are smaller in volume but larger in amplitude of
$|{\delta \rho \over \rho}|$.

As noted above, the smallness of $|S_3|$ is due to the fact that
the region where $|\Phi^{(2)}|$ along a photon trajectory becomes
comparable to $|\phi_0|$ ($z\sim 1,~\eta \sim \eta_0$)
is widely spatially separated from the recombination surface
$\eta =\eta_{rec}$ where $\left({\Delta T\over T}\right)_{loc}$
is located. Thus, to check that there is no other significant
contribution to $|S_3|$, we consider sixth-order terms that need
not possess this property.
The most important of them is the following term:
\begin{equation}
C_3^{(6)}(0)=\langle \left({\Delta T\over T}\right)_
{non-loc}^3\rangle~,  \label{sixth}
\end{equation}
where $\left({\Delta T\over T}\right)_{non-loc}$ is calculated with
the use of $\Phi^{(2)}$. The main contribution to (\ref{sixth})
is mainly produced at
recent times $\eta \sim \eta_0$ and it is not attenuated by a small
value of the spatial correlation function. However, it contains one
more power of $B$ as compared to (\ref{C3}).
The term (\ref{sixth}) can be represented as
\begin{eqnarray}
C_3^{(6)}(0)=-{16\eta_0^4\over 42^3} {8\over (2\pi)^7}\int \int \int
d^3k_1d^3k_2d^3k_3\, \phi^2(k_1)\phi^2(k_2)\phi^2(k_3)
\delta(k_{1z} - k_{2z})\delta(k_{2z} - k_{3z}) \nonumber \\
\times M({\bf k}_1,-{\bf k}_2)M({\bf k}_2,-{\bf k}_3)M({\bf k}_3,
-{\bf k}_1)  \label{int}
\end{eqnarray}
where the kernel $M({\bf k}_1,{\bf k}_2)$ is defined in Eq. (\ref{DMdef})
below (see Appendix B for a detailed derivation). Numerical evalulation
of this integral gives
\begin{equation}
C_3^{(6)}(0)= -2.15\times 10^6~B^3, ~~~  S_3^{(6)}~\approx -10^{-3}~.
\end{equation}
Thus, the sixth order contribution is much  smaller than the fourth
order one.

\section {Conclusions}

We have calculated the mean CMB skewness generated due
to leading non-linear corrections to an initially Gaussian
adiabatic perturbations with the flat spectrum. The unsmoothed value
$S_3(0) \approx -2$ that we found lies much below the cosmic variance
of this quantity $\delta S_3 \approx (\sigma_Tl_c)^{-1}\approx
130$ (here we model $C_2(\vartheta)$ by the Gaussian
$C_2=\sigma_T^2\exp \left(-{(l_c\vartheta)^2\over 2}\right)$
with $\sigma_T=3\cdot 10^{-5}$ and $l_c=250$ corresponding to the first
Doppler peak). The situation for smoothed large-angle maps is even
worse (e.g., $S_3\approx -5$ but $\delta S_3
\approx 3600$ for $\theta_{FWHP}=10^{\circ}$). Therefore,
as regarding observations, the prediction is that there should be no
noticeable mean skewness above the noise level due to cosmic variance
in the case of the standard CDM model with the flat (Harrison-Zeldovich)
initial spectrum of adiabatic perturbations.
This conclusion seems to be in a good agreement with existing
data ( Hinshaw et al. 1994, Kogut et al. 1994). It may
be considered as one more confirmation of predictions of the
inflationary scenario, though, of course, this fact does not close
the way for other theories leading to the same prediction.

It is clear from our derivation that $|S_3|$ can be substantially
larger if there exists a first order contribution to
$\left({\Delta T\over T}\right)_{non-loc}$ at late times ($z\sim 1$),
because then correlations
between first and second order terms in $C_3(0)$ are not small.
It may happen, as mentioned above, for the flat CDM$+\Lambda$
cosmological model and other ones. We shall consider this case
elsewhere.

\acknowledgements

A.S. is grateful to Profs. Y. Nagaoka and J. Yokoyama for their
hospitality at the Yukawa Institute for Theoretical Physics,
Kyoto University. A.S. was supported in part by the Russian
Foundation for Basic Research, Project Code 93-02-3631, and by
Russian Research Project ``Cosmomicrophysics''. It is a pleasure for
D.M. and T.S. to thank their supervisor Dr. Varun Sahni for his
encouragement and active support during the course of this work.
D.M. and T.S. also thank S. Bharadwaj for very useful discussions.
D.M. was financially supported by the Council of Scientific and Industrial
Research, India, under its SRF scheme.

\appendix
\section{ Fourth order contribution}

We outline some of the details of our estimation of the skewness.
The lowest order term leading to skewness  (4th. order in $\phi_0$)
is given by

\begin{eqnarray}
C_3^{(4)}(0) =  {2 \over 63} \int^{\eta_0}_{\eta_{rec}}
d\eta~ {\partial \over \partial \eta}\langle \phi_0^2({\bf r}_1)
\Phi^{(2)}({\bf  r}_2)\rangle~,~~~
|{\bf r}_1| = r_0,~~{\bf r}_2 = {\bf r}_1{\eta_0 - \eta \over r_0},
\label{A1}
\end{eqnarray}

\noindent
where $\phi_0$ and $\Phi^{(2)}$ represent values of potential
fluctuations at the linear and second order, respectively,
and $\Phi^{(2)}$ is regularized according to Eq. (\ref{ren});
$r_0$ is defined in Eq. (\ref{SW}). In
the paper, we follow the notation $\phi = \phi_U - \phi_M - \phi_D$
and $\Phi^{(2)} = \Phi_U - \Phi_M - \Phi_D$, where the subscripts
$U$, $M$ and $D$ represent unsubtracted, monopole and dipole fields,
correspondingly.

The CMB mean skewness at the lowest order can be then expressed
(see Eq. (\ref{sub})) as

\begin{equation}
C_3^{(4)}(0) = I_{UUU}-I_{UUM} -I_{UUD} - 2I_{UMU}+2I_{UMM}+I_{MMU}
-I_{MMM} + 2 I_{UMD} - I_{MMD} \label{A2}
\end{equation}

\noindent
where we have ignored terms involving $\phi_D$ since they are expected
to be small. In addition, we also expect that $I_{UMD}$ and $I_{MMD}$
are smaller in magnitude.

All the terms contain an average of the type $\langle \phi_0({\bf r}_1)
\phi_0({\bf r}_1') \Phi^{(2)}({\bf r}_2)\rangle$  which can be expressed
in terms of the potential-potential correlation function $\xi_\phi$
and its first derivative $\xi'_\phi$ as

$$\langle \phi_0({\bf r}_1) \phi_0({\bf r}_1') \Phi^{(2)}({\bf r}_2)
\rangle  =2 \nabla \xi_\phi( {\bf r}_1 - {\bf r}_2) \nabla
\xi_\phi(  {\bf r}_1' - {\bf r}_2)+ 5 \Delta^{-1}~[~ 2 \xi_\phi'({\bf r}_1
- {\bf r}_3)  \xi_\phi'( {\bf r}_1' - {\bf r}_3)$$
$$+ \nabla \xi_\phi({\bf r}_1 - {\bf r}_3) \nabla \xi_\phi'({\bf r}_1'
- {\bf r}_3)  + \nabla \xi_\phi'({\bf r}_1 - {\bf r}_3) \nabla
\xi_\phi({\bf r}_1' - {\bf r}_3)~]~. $$

The  terms $I_{UUM}$ and $I_{UUD}$ involve a simpler expression
with $ {\bf r}_1 \equiv {\bf r}_1'$
and can be easily evaluated in the coordinate space. We find that
the terms  $I_{UUM}$, $I_{UUD}$, $I_{UMU}$ and  $I_{UMM}$ are all
insensitive to the form of the transfer function $c(k)$  and may be
evaluated in the approximation $c(k) \equiv 1$, $\eta_{rec}=0$.

The term $I_{UUM}$ and $I_{UUD}$  involve the monpole and dipole term
of $\Phi^{(2)}$, respectively  and can be expressed as
\begin{eqnarray}
I_{UUM} ={2 \over 63}~{1 \over 4\pi}~ \int d\Omega_2 \int^{\eta_0}_
{{\eta}_{rec}} d\eta {\partial \over \partial \eta}\langle
\phi_0^2({\bf r}_1) \Phi^{(2)}({\bf r}_2)\rangle
\Big|_{|{\bf r}_1| = r_0,~ |{\bf r}_2| = \eta_0 - \eta}~,
\nonumber \\
I_{UUD} ={2 \over 63}~{3 \over 4\pi}~ \int d\Omega_2 \cos\theta_2
\int^{\eta_0}_{{\eta}_{rec}} d\eta {\partial \over \partial \eta}
\langle \phi_0^2({\bf r}_1) \Phi^{(2)}({\bf r}_2)\rangle
\Big|_{|{\bf r}_1| = r_0,~ |{\bf r}_2| = \eta_0 - \eta}~.
\end{eqnarray}

Assuming a scale invariant spectrum and $c(k) \equiv 1,~\eta_{rec}=0$,
we set $\xi_{\phi}(r)\approx \xi_{\phi}(0)-B\ln(r/r_{eq})$  to reduce
the expressions to
\begin{eqnarray}
I_{UUM}  = -{2 B^2 \over 21}{1\over 4\pi}\int_0^{\eta_0}
d\eta ~\eta \int {d\Omega \over r^2}~, \nonumber \\
I_{UUD}= -{2 B^2\over 21}{3\over 4\pi} \int_0^{\eta_0}d\eta~\eta
\int {\cos \theta \over r^2}~d\Omega
\end{eqnarray}

\noindent
where $r^2=(\eta_0-\eta)^2+\eta_0^2 - 2\eta_0(\eta_0-\eta)\cos \theta$.

Performing the integration over angles, we get
\begin {eqnarray}
I_{UUM} = - {B^2 \over 21}\int_0^1 {x\, dx\over 1 -x}\ln \left({2-x
\over x}\right) = -{2B^2 \over 21}({\pi^2 \over 8} - \ln 2)
\approx -0.0515B^2~, \nonumber \\
I_{UUD}=-{B^2\over 7}~\int_0^1 {x\, dx\over (1-x)^2}
\left[{(1-x)^2+1\over 2}~\ln \left({2-x\over x}\right)-1+x\right]=
-{B^2\over 7} \left( {3\over 2}-{\pi^2 \over 8}\right) \approx -0.0380B^2.
\end{eqnarray}

Now we consider the terms $I_{UMU}$ and $I_{UMM}$ which  involve the
monopole term of $\phi_0$ and  can be written as
\begin{eqnarray}
I_{UMU} ={2 \over 63}~{1 \over 4\pi}~ \int d\Omega_1' \int^{\eta_0}_
{{\eta}_{rec}} d\eta {\partial \over \partial \eta}\langle \phi_0
({\bf r}_1) \phi_0({\bf r}_1') \Phi^{(2)}({\bf r}_2)\rangle~, \nonumber \\
|{\bf r}_1| = |{\bf r}_1'|= r_0,~~ {\bf r}_2 = {\bf r}_1~
{\eta_0 - \eta \over r_0}~; \nonumber \\
I_{UMM} ={2 \over 63}~{1 \over 16\pi^2}~ \int d\Omega_1' \int
d\Omega_2 \int^{\eta_0}_{{\eta}_{rec}} d\eta {\partial \over
\partial \eta}\langle \phi_0({\bf r}_1) \phi_0({\bf r}_1')
\Phi^{(2)}({\bf r}_2)\rangle ~, \nonumber \\
|{\bf r}_1| = |{\bf r}_1'| = r_0,~ |{\bf r}_2| = \eta_0 - \eta ~.
\end{eqnarray}

It is more convenient to evaluate these terms in the momentum space.
In the $k$ - space, the unsubtracted, monopole and dipole terms for the
temperature fluctuation along the line of sight $ {\bf n}$ can be
expressed as

\begin {eqnarray}
\bigg({\Delta T \over T}\bigg)_U^{(1)} &=& {1 \over 3}\int {d^3k
\over (2\pi)^3} \phi_0({\bf k})e^{i{\bf kn} r_0 }~,\nonumber \\
\bigg({\Delta T \over T}\bigg)_M^{(1)} &=& {1 \over 3}
\int {d^3k \over (2\pi)^3} \phi_0({\bf k})~j_0(kr_0)~,\nonumber \\
\bigg({\Delta T \over T}\bigg)_D^{(1)} &=&  {i}\int {d^3k
\over (2\pi)^3} \phi_0({\bf k}) ~{{\bf kn}\over k}~j_1(kr_0)~,\nonumber \\
\bigg({\Delta T \over T}\bigg)_U^{(2)} &=&  {2 \over 21}
\int {d^3k \over (2\pi)^3} \Phi^{(2)}({\bf k})
\int_{\eta_{rec}}^{\eta_{0}}d\eta~\eta e^{i{\bf kn}( \eta_0 -\eta) }~,
 \nonumber \\
\bigg({\Delta T \over T}\bigg)_M^{(2)} &=&  {2 \over 21}\int {d^3k
\over (2\pi)^3} \Phi^{(2)}({\bf k})\int^{\eta_0}_{\eta_{rec}}
d\eta ~ \eta ~j_0(k(\eta_0-\eta))~,\nonumber \\
\bigg({\Delta T \over T}\bigg)_D^{(2)} &=&  {2i \over 7}
\int {d^3k \over (2\pi)^3} \Phi^{(2)}({\bf k})~ {{\bf kn}
\over k} ~\int^{\eta_0}_{\eta_{rec}}d\eta ~ \eta ~j_1(k(\eta_0-\eta))
\end{eqnarray}
where
$${d^3k \over (2\pi)^3} \Phi^{(2)}({\bf k}) =
-{d^3k_1 \over (2\pi)^3}{d^3k_2 \over (2\pi)^3}~M({\bf k}_1,
{\bf k}_2)~\phi_0({\bf k}_1) \phi_0({\bf k}_2),$$
$$M({\bf k}_1, {\bf k}_2) = 5 P({\bf k}_1, {\bf k}_2) +
Q({\bf k}_1, {\bf k}_2),$$
\begin {equation}
P({\bf k}_1, {\bf k}_2) = {2 k_1^2k_2^2+{\bf k}_1{\bf k}_2(k_1^2+k_2^2)
\over 2({\bf k}_1 +{\bf k}_2)^2},~~
Q({\bf k}_1, {\bf k}_2) ={\bf k}_1{\bf k}_2 ~, ~~ k \equiv |{\bf k}|~,
\label{DMdef}
\end{equation}

\noindent
the function $j_n(x)$ represents  the  standard spherical Bessel function
of order $n$. Note that  within the scope of this paper the first
order temperature fluctuation $(\Delta T/T)^{(1)} \equiv (\Delta
T/T)_{loc}$ and the second order temperature fluctuation $(\Delta
T/T)^{(2)} \equiv (\Delta T/T)_{non-loc}$, owing to the fact that the
linear gravitational potential does not change in a dust-dominated,
flat FRW universe.

In evaluating expectation values in the $k$-space we invoke the Gaussian
nature of the initial potential perturbations $\phi_0$ and implement
the following relation

$$ \langle \phi_0({\bf k}_1)\phi_0({\bf k}_2)\Phi^{(2)}({\bf k})
\rangle \rightarrow \langle \phi_0({\bf k}_1)\phi_0({\bf k}_2)
\phi_0({\bf k}_3)\phi_0({\bf k}_4)\rangle  = \phi^2(k_1)
\phi^2(k_2) $$
$$ \times \left( \delta({\bf k}_1 + {\bf k}_3)~\delta ({\bf k}_2
+ {\bf k}_4) + \delta ({\bf k}_1 + {\bf k}_4)~\delta ({\bf k}_2 +
{\bf k}_3) \right) ~, $$
\begin{equation}
\phi^2(k)={2\pi^2 B c^2(k)\over k^3}
\label {gauss2cdn}
\end{equation}
\noindent
where we have ignored the self-coupling term $\phi^2(k_1)~\delta
({\bf k}_1 +{\bf k}_2) \langle \Phi^{(2)}({\bf k})\rangle$
because it is just cancelled after the renormalization (\ref{ren}).

The expressions for the two terms now read

\begin{eqnarray}
I_{UMU} = -{ B^2\over 63}\int_{\eta_{rec}}^{\eta_0} d\eta \, \eta
\int {d^3k_1 \over 2\pi}\int {d^3k_2 \over 2\pi}~{c^2(k_1)
\over k_1^3}~{c^2(k_2) \over k_2^3} M({\bf k}_1,{\bf k}_2)
e^{-i{\bf k}_1{\bf n} (\eta - \eta_{rec})} e^{i{\bf k}_2{\bf n}
(\eta_0-\eta)}j_0( k_2r_0)~,
\nonumber \\
I_{UMM} = -{ B^2\over 63}\int_{\eta_{rec}}^{\eta_0} d\eta \, \eta
\int {d^3k_1
\over 2\pi}\int {d^3k_2\over 2\pi}~{c^2(k_1) \over k_1^3}~
{c^2(k_2) \over k_2^3} M({\bf k}_1,{\bf k}_2)~e^{i({\bf k}_1+{\bf k}_2)
{\bf n}(\eta_0-\eta)}~j_0( k_1r_0)~j_0( k_2r_0). \label{umuumm}
\end{eqnarray}

We find it convenient to split each of the terms into two pieces
involving the local part $Q$ and the non-local part $P$ part of
$\Phi^{(2)}$, respectively. After carrying out some of the integrations
 analytically, the terms can be expressed as

$$I_{{UMU}_Q}= -{4 B^2 \over 63} \int^{\eta_0}_{\eta_{rec}} d\eta~\eta
\int_0^{\infty} dk_1~c^2(k_1)~j_1(k_1 (\eta - \eta_{rec}))~
\int_0^{\infty} dk_2 c^2(k_2)~j_0(k_2r_0)~j_1( k_2 (\eta_0-\eta))~, $$

$$I_{{UMU}_P} = -{5B^2 \over 63} \int_{\eta_{rec}}^{\eta_0} d\eta~
\eta~\int_0^{\infty} dk_1~c^2(k_1)~\int_0^{\infty} dk_2~c^2(k_2)~
j_0(k_2r_0)$$
$$\times \int_{-1}^1 du \bigg[ {( k_1^2 + k_2^2)u + 2k_1k_2
\over k_1^2 + k_2^2 + 2k_1 k_2 u }\bigg]~j_0 \left( \sqrt{ k_1^2
(\eta-\eta_{rec})^2+ k_2^2(\eta_0 - \eta)^2 - 2 k_1k_2 (\eta - \eta_{rec})
( \eta_0 - \eta) u} \right)~,$$

$$ I_{{UMM}_Q} = {4 B^2 \over 63} \int^{\eta_0}_{\eta_{rec}} d\eta~\eta
\bigg[ \int dk~c^2(k)~j_0(k r_0)~j_1(k(\eta_0 - \eta)) ~\bigg]^2~,$$

$$ I_{{UMM}_P} = - {5B^2 \over 63}\int^{\eta_0}_{\eta_{rec}} d\eta~\eta
\int dk_1 c^2(k_1)j_0(k_1 r_0)~\int dk_2 c^2(k_2)j_0(k_2 r_0)$$
\begin{equation}
\times{\int_{-1}^1 du }~ \bigg[{{2k_1k_2 + (k_1^2 + k_2^2)u}
\over {k_1^2 + k_2 ^2 + 2k_1k_2u}}\bigg] ~~j_0\left( (\eta_0 - \eta)
\sqrt{k_1^2 + k_2^2 + 2k_1k_2 u}\right)~.
\end{equation}

The expressions are further simplified if we take $c(k) \equiv 1$ and
$\eta_{rec} = 0$ to obtain the final results

$$I_{{UMU}_Q} = -{2B^2\over 63} \int^1_0 {dx \over 1 -x}~\bigg[ 1 +
{ x (x - 2) \over 1 -x }~{\rm tanh}^{-1}(1 -x)\bigg] =-{2B^2 \over 63}~
(2 \ln 2 -1) \approx  -0.0123B^2~,$$
\begin{equation}
I_{{UMU}_P} = -0.23B^2, ~~I_{UMU} = I_{{UMU}_Q} + I_{{UMU}_P} = -0.24B^2
\end{equation}

\noindent and

$$I_{{UMM}_Q} = {B^2 \over 63} \int^1_0 {x~dx \over {( 1 - x )^2}}~~
\bigg[ 1 + {x(x - 2)\over  1 - x }~~{\rm tanh}^{-1} ( 1 - x) \bigg]^2
\approx 7.30 \times 10^{-4} B^2~,$$
\begin{equation}
I_{{UMM}_P}\approx -0.12B^2, ~~~~I_{UMM} = I_{{UMM}_Q} + I_{{UMM}_P} =
-0.12B^2~.
\end{equation}

\noindent We have numerically verified the above terms including the CDM
transfer function and value of $\eta_{rec}$.

Now we deal with the most significant term $I_{UUU}$ which depends
sensitively on small-scale power in the radiation perturbation spectrum
and transfer functions for radiation (see below). As noted before,
for the angular range $5'< \vartheta < 2^{\circ}$ that roughly
corresponds to the scale range $10$ Mpc $< a(\eta_0)(kh)^{-1}
< 200$ Mpc, all three effects - the Sachs-Wolfe,
the Doppler and the Silk ones - should be taken into account in
the local term $(\Delta T/T)_{loc}$. Moreover, only the
sum of the three effects appears to
be a gauge-invariant quantity. On the other hand, the recombination
may be still considered as instanteneous (the width of the
recombination surface in terms of the conformal time $\Delta \eta_{rec}
\ll k^{-1}$). More exactly, the effect of the non-zero width
$\Delta \eta_{rec}$ may be empirically accounted by the increase of
the scale of Silk damping. Therefore, we use (partly unpublished)
results of Starobinsky \& Sahni 1984 and Sahni 1984 (see also
Starobinsky 1987, 1988) obtained in the two-fluid approximation of the CDM
model with radiation. Namely,
the limiting case $\Omega_b=0$ was first considered and the matter
before recombination was assumed to consist of two ideal fluids
interacting through gravity only:
dust with pressure $p=0$ representing cold dark matter and radiation
with $p=\epsilon/3$ representing photons (tightly coupled with baryons)
and other massless particles.
After the recombination, photons are described as free massless
particles. This approach is equivalent to that used in Seljak 1994.
Also, we assume that $\eta_{eq}\ll \eta_{rec}$. However,
a small, but non-zero value of $\Omega_b$ should be finally taken
into account because it is crucial for the determination of the
Silk damping scale. Also, we accounted for it in the value
of the sound velocity in the radiation component before recombination
to get the right location of the acoustic peaks.

We use the following values in actual caculations:
$H_0=50$ km/s/Mpc, $T=2.726$ K, $\kappa = 1.681, ~\Omega_{tot}=1,
{}~\Omega_b=0.06,~z_{rec}=1100$. The scale factor of the two-fluid
(dust and radiation) CDM model has the form $a(\eta)=a_1\eta (\eta+
\eta_1)$ where $\eta_1=2(\sqrt{2} +1 )\eta_{eq}=4.828\eta_{eq}$.
Now we can find values of $\eta_{eq},~\eta_1$ and $\eta_{rec}$ using
the present-day value of $\Omega_{\gamma}=0.9895\times 10^{-4}$:
\begin{eqnarray}
{\eta_0 \over \eta_1}\approx (2\sqrt{\kappa \Omega_{\gamma}})^{-1}
(1-\sqrt{\kappa \Omega_{\gamma}})=38.3~, ~~{\eta_0\over \eta_{eq}}=
185~,~~z_{eq}=(\kappa \Omega_{\gamma})^{-1}=6012~, \nonumber \\
{\eta_{rec}\over \eta_1}={-1+\sqrt{1+(\kappa \Omega_{\gamma}
z_{rec})^{-1}}\over 2}=0.771~, ~~{\eta_0\over \eta_{rec}}=49.6~.
\label{eta}
\end{eqnarray}
It is clear that the approximation $\eta_{eq}\ll \eta_{rec}$ has
$\approx 15\%$ accuracy.

Then it can be shown (see, e.g. Starobinsky 1988) that, in the
approximation outlined above, $(\Delta T/T)_{loc}$ has the
following structure:
\begin{eqnarray}
\left( {\Delta T\over T}\right)_{loc}({\bf k}) &=& {1\over 3}
\tilde \phi_0({\bf k})
\left( F(k)+3ic_s{{\bf kn}\over k}G(k)\right) \exp
\left(-{k^2R_S^2\over 2}\right)~, \nonumber \\
F(k) &=& f(k)\cos (kc_s\eta_{rec})-g(k)\sin (kc_s\eta_{rec})~, \nonumber \\
G(k) &=& f(k)\sin (kc_s\eta_{rec})+g(k)\cos (kc_s\eta_{rec})~, \nonumber \\
c_s &=& {1\over \sqrt{3(1+R)}} =0.486~, ~~R\equiv \left({3\Omega_b
\over 4\Omega_{\gamma}}\right)_{rec}=0.413  \label{FG}
\end{eqnarray}
where $\tilde \phi_0 ({\bf k})$ is the {\em initial} perturbation
spectrum, i.e. $\tilde \phi_0 ({\bf k})= \phi_0 ({\bf k})/c(k)$,
$c_s$ is the sound velocity in the radiation component coupled
with baryons (in terms of the light velocity) and $R_S$ is the
Silk damping scale. Following Bond 1988,
we assume $R_S=12.4$ Mpc (for the values of $H_0$ and $\Omega_b$
given above). This value of $R_S$ partly accounts for an increase
of the Silk damping scale during recombination, and the final result
for temperature fluctuations appears to be in a good agreement with
numerical calculations using the exact kinetic approach (Bond \& Efstathiou
1987, Scott \& White 1994, Sugiyama 1994). The dependence of
the transfer functions $f(k)$ and $g(k)$ on $\Omega_b$ is weak
and we neglect it. Their form in the limit
$\Omega_b \to 0$ calculated in Sahni 1984 may be well approximated
by the following analytical fits in terms of the variable
$x=k\eta_1=ka^{-1}(\eta_0)\times 309$ Mpc:
\begin{eqnarray}
f(x) &=& 1+0.042x^2 ~~~~~~~~~x \le 1.55 \nonumber \\
     &=& 0.8991+0.1302x~~~~~~1.55 < x \le 9.379  \nonumber \\
     &=& {5x\over x+12.74}~~~x>9.379~,  \nonumber \\
g(x) &=& {3\ln \left( 1+{x\over 10}\right) \over 1+{x\over 16}
+{3x^2\over 400}}~. \label{fg}
\end{eqnarray}
The oscillating  terms in (\ref{FG}) are the well-known acoustic
oscillations in the baryon-photon plasma (see, e.g., Zeldovich \& Novikov
1983 and references therein).

Substituting (\ref{FG}) into Eq. (\ref{four}), we get after some
algebra:
$$I_{UUU}=-{4\over 63}\int_{\eta_{rec}}^{\eta_0} d\eta~\eta~
{1\over (2\pi)^6}\int {d^3k_1 \over k_1^3} \int {d^3k_2\over k_2^3}~
M({\bf k}_1,{\bf k}_2)~e^{-i({\bf k}_1+{\bf k}_2){\bf n}(\eta -
\eta_{rec})} $$
$$\times (2\pi^2B)^2c(k_1)c(k_2)~e^{ -{1\over 2 }(k_1^2+k_2^2) R_S^2}
\left( F(k_1)-3ic_s{{\bf k}_1{\bf n}
\over k_1}G(k_1)\right) \left( F(k_2)-3ic_s{{\bf k}_2{\bf n}
\over k_2}G(k_2)\right) $$
$$= -{B^2\over 63}\int_{\eta_{rec}}^{\eta_0}d\eta~\eta~\int_0^{\infty}
dk_1\int_0^{\infty}dk_2\int_{-1}^1du~\left( 2u+{5(2k_1k_2+(k_1^2
+k_2^2)u)\over k_1^2+k_2^2+2k_1k_2u}\right)~c(k_1)c(k_2)
{}~e^{ -{1\over 2 }(k_1^2+k_2^2) R_S^2}$$
$$\times \lbrack F(k_1)F(k_2)j_0(w)-3c_s\left( F(k_1)G(k_2)(k_1u+k_2)
+G(k_1)F(k_2)(k_1+k_2u)\right){j_1(w)(\eta-\eta_{rec})\over w} $$
$$+ 9c_s^2G(k_1)G(k_2)\left( {j_2(w)(\eta-\eta_{rec})^2(k_1+k_2u)
(k_1u+k_2)\over w^2}-{j_1(w)u\over w}\right)\rbrack~,$$
\begin{equation}
w=\sqrt{k_1^2+k_2^2+2k_1k_2u}~(\eta-\eta_{rec})~. \label{corint}
\end{equation}
Calculating this integral numerically, we obtain
\begin{equation}
I_{UUU}  \approx  -2.2~B^2.
\end{equation}
Note for comparison that if we did not take into account the increase
of $({\Delta T\over T})_{loc}$ due to the Silk and Doppler effects, i.e.
if we calculated the integral (\ref{corint}) in the limit $f(k)\equiv 1,~
g(k)\equiv 0$ (but with the exact $c(k)$) we would get the answer
$I_{UUU}= -0.9~B^2$. So, the account of all effects increase the answer
by a factor of $2.4$.

\setcounter{equation}{0}
\section{ Sixth order contribution}

Beyond the fourth order term, the next contribution to the skewness
arises at the sixth order in  $\phi_0$. In this section,  we outline
the calculation for the most significant of the  sixth order terms,
$\langle ({\Delta T/ T})^{(2)}_U(\Delta T/ T)^{(2)}_
U({\Delta T/ T})^{(2)}_U\rangle$. Substituting the expression for
$({\Delta T/ T})^{(2)}_U$ from (\ref{DMdef}) and (\ref{gauss2cdn})
we obtain

$$ C^{(6)}_3 (0) = -{64\over (42)^3}~\int_{ \eta _{rec}}^{\eta _0}
 d \eta_1 \eta_1 \int {d^3k_1 \over (2\pi)^3} \int {d^3k_1' \over (2\pi)^3}
 [5 P({\bf k}_1, {\bf k}_1')+ Q({\bf k}_1, {\bf k}_1')]
{\rm e}^{i~(k_1 \cos\theta_1 +k_1' \cos\theta_1')( \eta_0 - \eta_1)} $$
$$\times \int_{ \eta _{rec}}^{\eta _0}
 d \eta_2 \eta_2 \int {d^3k_2 \over (2\pi)^3} \int {d^3k_2' \over (2\pi)^3}
  [5 P({\bf k}_2, {\bf k}_2')+ Q({\bf k}_2, {\bf k}_2')]
{\rm e}^{i~(k_2 \cos\theta_2 + k_2' \cos\theta_2')( \eta_0 - \eta_2)}$$
$$\times \int_{ \eta _{rec}}^{\eta _0}
 d \eta_3 \eta_3 \int {d^3k_3 \over (2\pi)^3} \int {d^3k_3' \over (2\pi)^3}
 [5 P({\bf k}_3, {\bf k}_3')+ Q({\bf k}_1, {\bf k}_3')]
{\rm e}^{i~(k_3 \cos\theta_3 + k_3' \cos\theta_3')( \eta_0 - \eta_3)}$$
\begin{equation}
\times \langle \phi ({\bf k_1}) \phi ({\bf k}_1') \phi ({\bf k}_2)
\phi ({\bf k}_2') \phi ({\bf k}_3) \phi ({\bf k}_3') \rangle, \label{sixth1}
\end{equation}

Assuming the initial potential fluctuations to be a gaussian random
field (as predicted by most inflationary scenarios) we have

$$ \langle \phi ({\bf k}_1) \phi ({\bf k}_1') \phi ({\bf k}_2)
\phi ({\bf k}_2') \phi ({ \bf k}_3) \phi ({\bf k}_3')
\rangle = \phi^2(k_1)\phi^2(k_2)\phi^2(k_3)~ $$
\begin{equation}
\times \left( \delta({ \bf k}_1 + {\bf k}_2') ~\delta({\bf k}_2 +
{\bf k}_3')~ \delta({ \bf k}_3 + {\bf k}_1') ~~
+ 7~~ permutations \right) \label{gausscdn}
\end{equation}

\noindent  where we ignore the self-coupling terms. We assume
a scale-invariant initial spectrum\footnote{ Our calculations can be
trivially extended to non-scale-invariant spectra by incorporating
the appropiate form for
$\phi^2(k)$.} with $\phi^2(k)=2\pi^2Bc^2(k)/ k^3$,
where $B$ is a normalisation constant, and introduce the notations
$u \equiv \cos\theta_1$, $v \equiv \cos\theta_2$  and  $w \equiv
\cos\theta_3$ for brevity. The expressions can be simplified by making the
approximation

$$\int_{\eta_{rec}}^{\eta_0} \eta_1 d\eta_1 \int_{\eta_{rec}}^{\eta_0}
\eta_2 d\eta_2 \int_{\eta_{rec}}^{\eta_0} \eta_3 d\eta_3
e^{ik_1(\eta_2 - \eta_1)u}e^{ik_2(\eta_3 - \eta_2)v}
e^{ik_3(\eta_1 - \eta_3)w}$$
$$\approx \int_0^{\eta_0}\eta_1^3 \times {\int_{-\infty}^{\infty} dx}
{\int_{-\infty}^{\infty} dy}~e^{ik_1u x}e^{ik_2(y - x)v}
e^{-ik_3wy}$$

\begin{equation}
 = {\eta_0^4 \over 4}(2\pi)^2~\delta(k_1u - k_2v)~\delta(k_2v - k_3w)
= \pi^2 \eta_0^4~\delta(k_{1z} - k_{2z})~\delta(k_{2z} - k_{3z})
 \label {starapprx}
\end{equation}
\noindent
where $x=\eta_2-\eta_1,~y=\eta_3-\eta_1$.
Substituting (\ref{starapprx}) and (\ref{gausscdn}) into expression
(\ref{sixth1}), $C^{(6)}_3(0)$ can be written as (\ref{int}).

As in the case of the fourth order calculations, it is convenient to
split $C^{(6)}_3(0)$ into pieces involving the local and the non-local
parts of $\Phi^{(2)}$. We express

\begin{equation}
C^{(6)}_3(0) = {\cal I}_{QQQ} + 3 {\cal I}_{QQP} + 3{\cal I}_{QPP}
 + {\cal I}_{PPP}.
\label{split}
\end{equation}

Splitting the expression for  $C^{(6)}_3(0)$ as described by (\ref{split}),
we obtain the following results for the constituent terms:

\begin{eqnarray}
&{}&{\cal I}_{QQQ} = 4~\bigg({B \over 21}\bigg)^3
\int dk~k^3~c^2(k) {\int_0^{2 \pi} } d\alpha{\int_0^{2 \pi} } d\beta \int_0^1
 du \int_0^1 dv \int_0^1 dw~c^2({kv \over u})~c^2({kv \over w})~
{ v^2 \over u w} \nonumber \\
&\times& \cos\theta_1~\cos\theta_2~ \cos\theta_3 =  1.4\times 10^{6} B^3,
\label{qqq}
\end{eqnarray}

\bigskip

\begin{eqnarray}
&{}&{\cal I}_{QQP} = 10~\bigg({B \over 21}\bigg)^3
\int dk~k^3~c^2(k) {\int_0^{2 \pi} } d\alpha{\int_0^{2 \pi} } d\beta \int_0^1
 du \int_0^1 dv \int_0^1 dw~c^2({kv \over u})~c^2({kv \over w})~
{ v^2 \over u w} \nonumber \\
&\times& ~\left[{ r^2 \cos\theta_1 -2 u v \over r^2 - 2 u v \cos\theta_1}
\right]
{}~\cos\theta_2~ \cos\theta_3  =  - 8.1\times 10^{5} B^3,
\label{qqp}
\end{eqnarray}

\bigskip

\begin{eqnarray}
&{}&{\cal I}_{QPP} = 25~\bigg({B \over 21}\bigg)^3
\int dk~k^3~c^2(k) {\int_0^{2 \pi} } d\alpha{\int_0^{2 \pi} } d\beta \int_0^1
du \int_0^1 dv \int_0^1 dw~c^2({kv \over u})~c^2({kv \over w})
{}~{ v^2 \over u w} \nonumber \\
&\times& ~\left[{ r^2 \cos\theta_1 -2 u v \over r^2 - 2 u v \cos\theta_1}
\right]
{}~\left[{ s^2 \cos\theta_2  -2 w v \over s^2 - 2 w v \cos\theta_2}\right]
 \cos\theta_3  =  1.19\times 10^{7} B^3,
\label{qpp}
\end{eqnarray}

\bigskip

\begin{eqnarray}
&{}&{\cal I}_{PPP} = {1\over 2}~\bigg({5B \over 21}\bigg)^3
\int dk~k^3~c^2(k) {\int_0^{2 \pi} } d\alpha{\int_0^{2 \pi} } d\beta
\int_0^1 du \int_0^1 dv \int_0^1 dw~c^2({kv \over u})~c^2({kv \over w})
{}~{ v^2 \over u w} \nonumber \\
&\times& ~\left[{ r^2 \cos\theta_1 -2 u v \over r^2 - 2 u v \cos\theta_1}
\right]
{}~\left[{ s^2 \cos\theta_2  -2 w v \over s^2 - 2 w v \cos\theta_2}\right]
{}~\left[{ t^2 \cos\theta_3 -2 u w \over t^2 - 2 u w \cos\theta_3}\right] =
 -3.68\times 10^{7} B^3,
\label{ppp}
\end{eqnarray}

\noindent where we have used the following notations
\begin{eqnarray}
\cos\theta_1 &=& uv + \bar u \bar v \cos \alpha,~~\cos\theta_2 = uw +
\bar u \bar w \cos\beta, ~~ \cos\theta_3 = wv + \bar w \bar v \cos \beta,
\nonumber \\
r^2 &=& u^2 + v^2 ,~~ s^2 = v^2 + w^2,~~ t^2 = w^2 + u^2.
\label{notation}
\end{eqnarray}

\noindent We have evaluated the expressions (\ref{qqq}), (\ref{qqp}),
(\ref{qpp}) and (\ref{ppp}) numerically to estimate the contribution to
the skewness at the $6$th order in the potential fluctuation $\phi_0$.


\clearpage
\begin{thebibliography}{}
\bibitem{} Abbott L. F., \& Schaeffer R. K. 1986, ApJ, 308, 546.
\bibitem{} Bond, J. R. 1988, in: The Early Universe, ed. W. Unruh
\& G. Semenoff, Dordrecht (Reidel).
\bibitem{} Bond, J. R., \& Efstathiou, G. 1987, MNRAS, 226, 655.
\bibitem{} Falk, T., Rangarajan, R., \& Srednicki, M. 1993, ApJ, 403, L1.
\bibitem{} Gangui, A., Lucchin, F., Matarrese, S., \& Mollerach, S.
ApJ, 1994, 430, 447.
\bibitem{} Gorski, K. M., et al. 1994, ApJ, 430, L89
\bibitem{} Hinshaw, G., et al. 1994, ApJ, 431, 1.
\bibitem{} Kofman, L. A., \& Starobinsky, A. A. 1985, Sov. Astron.
Lett., 11, 271.
\bibitem{} Kofman, L. A., Pogosyan, D. Yu., \& Starobinsky, A. A. 1986,
Sov. Astron. Lett., 12, 175.
\bibitem{} Kogut, A., et al. 1994. COBE preprint 94-15, astro-ph/9408070.
\bibitem{} Luo, X., \& Schramm, D. N. 1993, Phys. Rev. Lett., 71, 1124.
\bibitem{} Martinez-Gonzalez, E., Sanz, J. L., \& Silk, J. 1992,
Phys. Rev. D, 46, 4193.
\bibitem{} Munshi, D., \& Starobinsky, A. A. 1994, ApJ, 428, 433.
\bibitem{} Peebles, P. J. E. 1980, The Large-Scale Structure of the
Universe (Princeton: Princeton University Press).
\bibitem{} Rees, M., \& Sciama, D. 1968, Nature, 217, 511.
\bibitem{} Sachs, R. K., \& Wolfe, A. M. 1967, ApJ, 147, 73.
\bibitem{} Sahni, V. 1984, Ph. D. thesis, Moscow University (unpublished).
\bibitem{} Scott, D. \& White, M. 1994, preprint astro-ph/9407073.
\bibitem{} Seljak, V. 1994, ApJ, 435, L87.
\bibitem{} Srednicki, M. 1993, ApJ, 416, L1.
\bibitem{} Starobinsky, A. A. 1983, Sov. Astron. Lett., 9, 302.
\bibitem{} Starobinsky, A. A. 1987, SAO Commun., 53, 57.
\bibitem{} Starobinsky, A. A. 1988, Sov. Astron. Lett., 14, 166.
\bibitem{} Starobinsky, A. A. \& Sahni, V. 1984, in: Proc. 6th Soviet
Gravitation Conf., ed. V. N. Ponomarev (MGPI Press, Moscow), p. 77.
\bibitem{} Sugiyama, N. 1994, Preprint CfPA-TH-94-62, astro-ph/9412025.
\bibitem{} Wilson M. L. 1983, ApJ, 273, 2.
\bibitem{} Zeldovich, Ya. B., \& Novikov, I. D. 1983, The Structure and
Evolution of the Universe (Chicago: Univ. Chicago Press).
\bibitem{} Zeldovich, Ya. B., \& Sazhin, M.V. 1987, Sov. Astron.
Lett., 13, 145.
\end{thebibliography}
\end{document}